\documentclass[conference]{IEEEtran}
\IEEEoverridecommandlockouts
\usepackage{cite}
\usepackage{amsmath,amssymb,amsfonts}
\usepackage{algorithmic}
\usepackage{graphicx}
\usepackage{textcomp}
\usepackage{xcolor}
\usepackage[markup=underlined]{changes}
\definechangesauthor[name={AK}, color=yellow]{Y}
\usepackage[inkscapelatex=false]{svg}
\usepackage{siunitx}
\usepackage[bookmarks=true,breaklinks=true,colorlinks,citecolor=blue,linkcolor=red,urlcolor=blue]{hyperref}
\def\BibTeX{{\rm B\kern-.05em{\sc i\kern-.025em b}\kern-.08em
    T\kern-.1667em\lower.7ex\hbox{E}\kern-.125emX}}

\newcommand{\s}{\phantom{0}}

\newcommand{\avgSpeedup}{1.56$\times$}
\newcommand{\maxSpeedup}{3.01$\times$}

\begin{document}

\title{Accelerating Detailed Routing Convergence through Offline Reinforcement Learning\vspace{-3mm}}


\author{\IEEEauthorblockN{Afsara Khan}
\IEEEauthorblockA{\textit{Electrical and Computer Engineering} \\
\textit{New York University}\\
Brooklyn, NY, USA \\
atk331@nyu.edu\vspace{-10mm}}
\and
\IEEEauthorblockN{Austin Rovinski}
\IEEEauthorblockA{\textit{Electrical and Computer Engineering} \\
\textit{New York University}\\
Brooklyn, NY, USA \\
rovinski@nyu.edu\vspace{-10mm}}
}

\maketitle

\begin{abstract}
Detailed routing remains one of the most complex and time-consuming steps in modern physical design due to the challenges posed by shrinking feature sizes and stricter design rules.
Prior detailed routers 
achieve state-of-the-art results by leveraging iterative pathfinding algorithms to route each net. However, runtimes are a major issue in detailed routers, as converging to a solution with zero design rule violations (DRVs) can be prohibitively expensive.


In this paper, we propose leveraging reinforcement learning (RL) to enable rapid convergence in detailed routing by learning from previous designs. We make the key observation that prior detailed routers \textit{statically} schedule the cost weights used in their routing algorithms, meaning they do not change in response to the design or technology. By training a conservative Q-learning (CQL) model to dynamically select the routing cost weights to minimize the number of algorithm iterations, we find that our work completes the ISPD19 benchmarks with \avgSpeedup{} average and up to \maxSpeedup{} faster runtime than the baseline router while maintaining or improving the DRV count in all cases. We also find that this learning shows signs of generalization across technologies, meaning that learning designs in one technology can translate to improved outcomes in other technologies.

\end{abstract}


\section{Introduction}

Detailed routing remains one of the most challenging aspects of modern physical design due to the increasing complexity of scaling feature sizes. As technology nodes advance, the number of design rules increases dramatically and leads to significantly increased runtime for detailed routing.

Prior work in detailed routing has leveraged techniques such as Lee's Algorithm~\cite{lee1961algorithm}, A* search~\cite{arnold}, line search~\cite{hightower1969solution}, and others to perform detailed routing. Most notably, Dr. CU~\cite{chen2020detailed} and OpenROAD~\cite{ajayi2019openroad-gomactech,ajayi2019openroad} are the leading open-source detailed routers capable of solving the ISPD18 and ISPD19 detailed routing benchmarks.
In both cases, the general routing algorithms are the same:
1) A pathfinding algorithm such as A* or Dijkstra's algorithm is used to find the minimum cost path for each routed net, 2) a design rule check (DRC) is run to evaluate any DRVs, 3) violation costs are added to the corresponding nodes on the grid, and 4) the route is ripped up and rerouted to avoid violations. An ``iteration'' is defined as the number of times the algorithm must ripup and reroute nets to achieve a DRV-free solution, which can become prohibitively expensive for high-density designs.

Our major insight is that prior work only performs \textit{static} costing of search weights, that is, the costs that are applied to violations do not change in response to the design. In the case of Dr. CU, the violation costs are static during the entire algorithm. For OpenROAD, the costs are adjusted based solely on the current ripup iteration number.



One of the great challenges in trying to improve over the baseline implementations is that altering the violation costing can frequently lead the detailed router towards worse solutions which take longer to converge or do not converge to 0 DRVs.
During our study, we find that random exploration of weights leads to a significantly higher number of worse solutions than better solutions, and weights which work well on one design may not translate to another design. Therefore, the challenge in improving weight selection relies on both 1) finding weights which enable the router to converge to 0 DRVs and 2) converging to 0 DRVs in fewer iterations than the baseline approach.

In this work, we propose using reinforcement learning (RL) to learn from prior design implementations and determine sequences of weights which will converge to 0 DRVs in the fewest number of iterations.
We use perturbation data sampling \cite{shi2023provably,perturbation} to explore 350-400 weight sequences per design in our training set, and then we train a conservative Q-learning (CQL) model \cite{cql} on these samples to predict the best weights for the next iteration in order to minimize iterations to convergence. We find that on the ISPD19 benchmark, our work converges in 5\% fewer iterations on average and up to 31\%. This translates to a runtime speedup of \avgSpeedup{} on average and up to \maxSpeedup{}.

One of our key results is that we find that not only are the current weight schedules suboptimal, but that the conventional wisdom for weight selection is \textit{dramatically different} from the weights that our RL model selects for improved results.
We make the following contributions in this work:

\begin{itemize}
\item  Development of a detailed routing training dataset using perturbation sampling and random exploration techniques
\item  Identification of critical state variables and reward functions suitable for RL models in detailed routing 
\item  Implementation and evaluation of an offline RL approach using the conservative Q-learning (CQL) model to predict optimal cost weights for detailed routing.
\item  Demonstration of robustness across diverse designs 
\item  Analysis of cost weights vs. routing performance to demonstrate that learned weights can achieve better performance by disregarding conventional practice.
\item We aim to open-source all source code for training and inference along with publication of this paper.
\end{itemize}

The rest of the paper is organized as follows.
Section~\ref{sec:RelatedWork} discusses related work,
Section~\ref{sec:DataGeneration} discusses our methodology for data generation, Section~\ref{sec:ModelArchitecture} discusses the model architecture and training, Section~\ref{sec:ExperimentalResults} discusses the experimental results, and Section~\ref{sec:Conclusion} provides our conclusions.

\section{Related Work}
\label{sec:RelatedWork}

\subsection{Detailed Routers}
The foundation of modern detailed routing traces back to Lee's maze routing algorithm \cite{lee1961algorithm}, a breadth-first search (BFS) that guaranteed minimum cost paths. Rapid notable improvements on this base algorithm included the A* search using heuristics to guide the search \cite{arnold,bidir}, along with techniques like line search~\cite{hightower1969solution} to speed up execution. Core strategies in modern routers also include the iterative rip-up and reroute 
and sometimes leverage multicommodity flow concepts \cite{han2015}. TritonRoute \cite{kahng2020tritonroute} utilizes the prior findings to employ an iterative A*-based search with partitions, enhanced by dynamic boundary adjustments. Similarly, Dr. CU~\cite{chen2020detailed} uses Dijkstra's algorithm coupled with a sparse grid-graph and partitioning to efficiently route nets. As newer technology nodes arrive, routing research has shifted focus towards techniques such as gridless pin access \cite{nieberg2011}, routing under complex patterning like SADP \cite{ding2017,liu2014}, minimum area-sensitive path search \cite{chang2013}, and ILP-based formulations \cite{han2015}.

In terms of academic routers, OpenROAD~\cite{ajayi2019openroad} provides state-of-the-art performance, achieving 0 DRVs on the ISPD18~\cite{mantik2018ispd} benchmark and 0 DRVs on all but one test case on ISPD19~\cite{liu2019ispd}. OpenROAD's router is derived from TritonRoute~~\cite{kahng2018tritonroute,kahng2020tritonroute}, but several enhancements have been made and it is actively maintained to continue improving detailed routing results.
In this work, we propose using dynamic weighting of costs which are learned from prior designs. Prior works use a static scheduling of weights which do not depend on the design. Additionally, we find that RL-generated weights tend to differ \textit{significantly} from the expert-curated weights in the baseline implementations.

\subsection{Machine Learning Techniques in Routing}


Several prior works have shown benefits from applying traditional machine learning to global and detailed routing. Many works have focused on congestion and DRV hotspot prediction at the routing level~\cite{10,11}, while others have focused on predicting and avoiding pin access violations~\cite{12}. These predictions typically required secondary mechanisms to influence routing behavior. More direct guidance has been explored via reinforcement learning. For example, Chen et al. \cite{13} propose an online RL framework that combines graph neural networks (GNNs) with PPO policy learning.

Our work differs significantly from prior works as we propose layering an RL model on top of the iterative search in order to reach DRV convergence faster. Our technique is applicable to any router which uses violation costing for its search, and our technique is agnostic to technology node. To the best of our knowledge, no prior work has used RL to learn across detailed routing iterations to improve convergence.

\section{Data Generation}
\label{sec:DataGeneration}




As the first step in training our RL model, we first select the data which represents the state of the design at the end of an iteration. We captured the following values:

\begin{itemize}
    \item \textbf{Input Weights}: the weights applied to violations during the current iteration. We use 4 distinct weights: \textbf{drcCost}: cost applied to potential DRVs. \textbf{markerCost}: penalty to nodes adjacent to DRVs. \textbf{fixedShapeCost}: cost applied nodes which violate fixed objects, such as macro pins or obstructions. \textbf{markerDecay}: the decay rate of DRV costs from previous routing iterations.
    \item \textbf{Partition DRVs}: the number of DRVs for each chip routing partition at the end of the iteration.
    \item \textbf{Maximum Partition DRV}: the maximum number of DRVs across all partitions at the end of the iteration.
    \item \textbf{Neighbor DRVs}: For each partition, the number of DRVs in the adjacent partitions
    \item \textbf{Total DRVs}: the total DRVs at the end of the iteration
    \item \textbf{Total Wirelength}: total wirelength in um at the end of the iteration
\end{itemize}

Several of the values are derived from the same source data, however we capture them individually in order to simplify model training and reduce model complexity.
More details on these parameters are discussed in prior work~\cite{kahng2020tritonroute}.

\begin{figure}
    \centering
    \includegraphics[width=0.8\columnwidth]{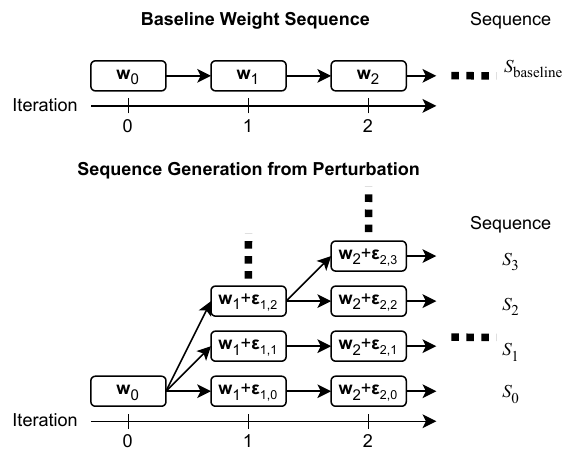}
    \caption{Top: the baseline implementation uses a single sequence of weight vectors ($\textbf{w}$). Bottom: we use perturbation sampling to create many new sequences per design. Sequences start near the baseline values and then gradually diverge with more samples (increasing $\varepsilon$).}
    \label{fig:perturbation}\vspace{-4mm}
\end{figure}

After each iteration, we sample each of these values as part of the state $s$. A ``sequence'' $S$ is formed by a series of states $\{s_1,s_2,...,s_n\}$ that result in DRV convergence, where $n$ is the number of iterations. Each sequence forms a data point which is used to train our RL model. Because states are deterministic, it is possible to represent a state simply by its sequence of input weight vectors. This is illustrated in Fig.~\ref{fig:perturbation} (top).

In addition to capturing the sequential history, we captured the following static design characteristics at the beginning of a design's routing flow to help guide the initial iteration: 
    die area,
    total number of pins,
    pin density,
    number of macros,
    instance density,
    number of nets,
    average pins per net,
    net density, and
    number of routing layers.
These characteristics are design-dependent and do not change during detailed routing.

One challenging aspect of data generation is generating useful sequences to obtain a dataset with enough sequences that quickly converge to zero DRVs. We employed a balanced approach between exploitation and exploration. For roughly 60\% of the dataset, we utilize perturbation-based sampling~\cite{perturbation}, beginning with weights near the baseline values established by domain experts and progressively exploring further from these values with each routing run. This process is illustrated in Fig.~\ref{fig:perturbation} (bottom). The remainder of the data set consisted of random sampling that leverage Sobol sequences~\cite{mishra2021enhancing,sobol} to ensure efficient coverage of the solution space while minimizing redundant data points.

Using this methodology, we conducted 5,782 detailed routing runs across 17 designs to generate a training set. The set included 10 designs from the OpenROAD Design Suite~\cite{rovinski2020bridging}, and 7 designs from the ISPD18 benchmarks to diversify technology nodes and design characteristics. ISPD19 circuits were excluded from training data to be used as test data.

\begin{figure}
    \centering
    \includegraphics[width=0.8\columnwidth]{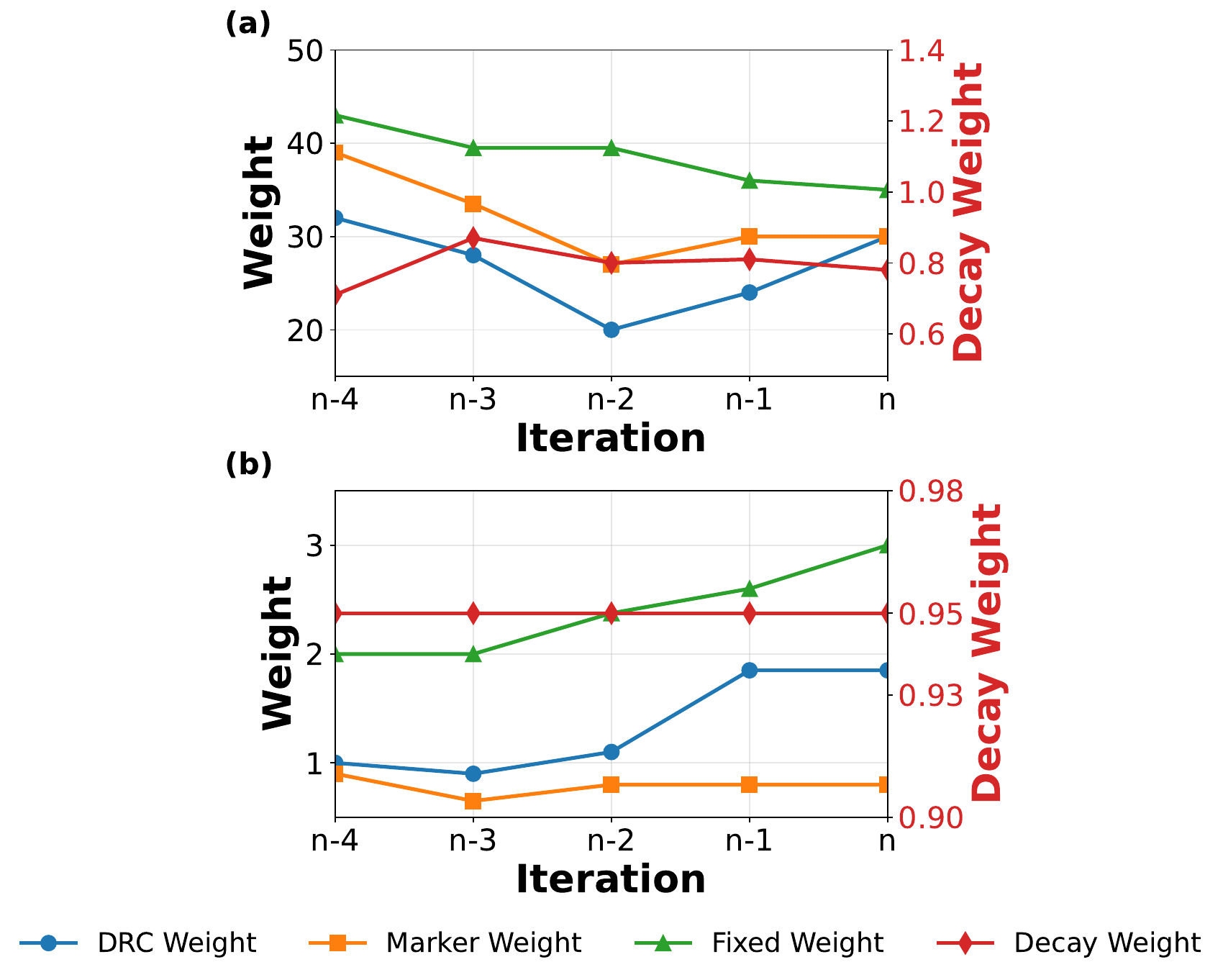}
    \caption{a) Optimal vs b) Baseline weight trend for the last few iterations across all designs in the dataset. Optimal weight trends differ substantially from the baseline.}
    \label{fig:average_weights}\vspace{-4mm}
\end{figure}

From this dataset, several trends become apparent. By taking the weight sequence for each design which resulted in the fastest convergence (minimum iterations), we observe how the weights differ from the baseline. Figure~\ref{fig:average_weights} (top) shows the average weights per iteration of these sequences. In this graph, $n$ represents the last iteration before convergence for the given design. Therefore, the graph shows the average weight over the last 5 iterations before convergence across all designs. For designs which converge in fewer than 5 iterations, only those weights are averaged. Figure~\ref{fig:average_weights} (bottom) shows the corresponding weights used in the baseline. We note that the optimal weights differ substantially from the baseline weights, both in the trends and the absolute value.
From these findings, we observe that there are several strong trends that can be exploited in order to improve detailed routing convergence. In the next section, we discuss our RL model architecture for inferring optimal weights.


\section{Model Architecture}
\label{sec:ModelArchitecture}

\subsection{Model}

Our architecture leverages Conservative Q-Learning (CQL), an offline reinforcement learning algorithm that addresses Q-learning's overestimation bias through value function regularization \cite{cql}.
CQL achieves this by incorporating a conservative penalty on out-of-distribution actions, effectively constraining the learned policy within the support of the training data.
We found online exploration impractical in this context, as each episode can take minutes to hours to complete, which is dramatically slower than the baseline.
The algorithm augments the standard Q-learning objective with a conservative term:
\begin{equation}
\begin{split}
\min_{Q} \Big( & \alpha\mathbb{E}_{s\sim\mathcal{D}}\big[\log\sum_{a}\exp(Q(s,a)) \\
& - \mathbb{E}_{a\sim\pi_\beta(a|s)}[Q(s,a)]\big] \\
& + \frac{1}{2}\mathbb{E}_{(s,a,r,s')\sim\mathcal{D}}\big[(Q(s,a) - \hat{\mathcal{B}}^\pi Q(s,a))^2\big] \Big)
\end{split}
\end{equation}
where $\alpha$ is the conservative penalty coefficient, $\mathcal{D}$ represents the offline dataset, and $\pi_\beta$ represents the behavior policy \cite{cql}. 
A key advantage of CQL for inferring routing weights is its reduced inference time, as the algorithm's conservative regularization inherently calibrates Q-values during training and eliminates the need for additional normalization or post-processing steps during deployment \cite{offlinerl2}.
Unlike conventional supervised learning approaches that require extensive datasets, CQL's architecture enables efficient learning from modest-sized datasets by leveraging complete state-action trajectories rather than isolated samples \cite{rlbook}. The model's ability to extract temporal dependencies and action-consequence relationships from each trajectory makes it particularly suitable for inferring routing weights as we attempt to optimize cost weights for every iteration by analyzing sequential trajectories from previous iterations. Additionally, CQL has a higher training stability due to prevention of Q-value overestimation in unexplored states while maintaining consistent performance across varying routing scenarios. 

\begin{figure}[t]
    \centering
    \includegraphics[width=\linewidth]{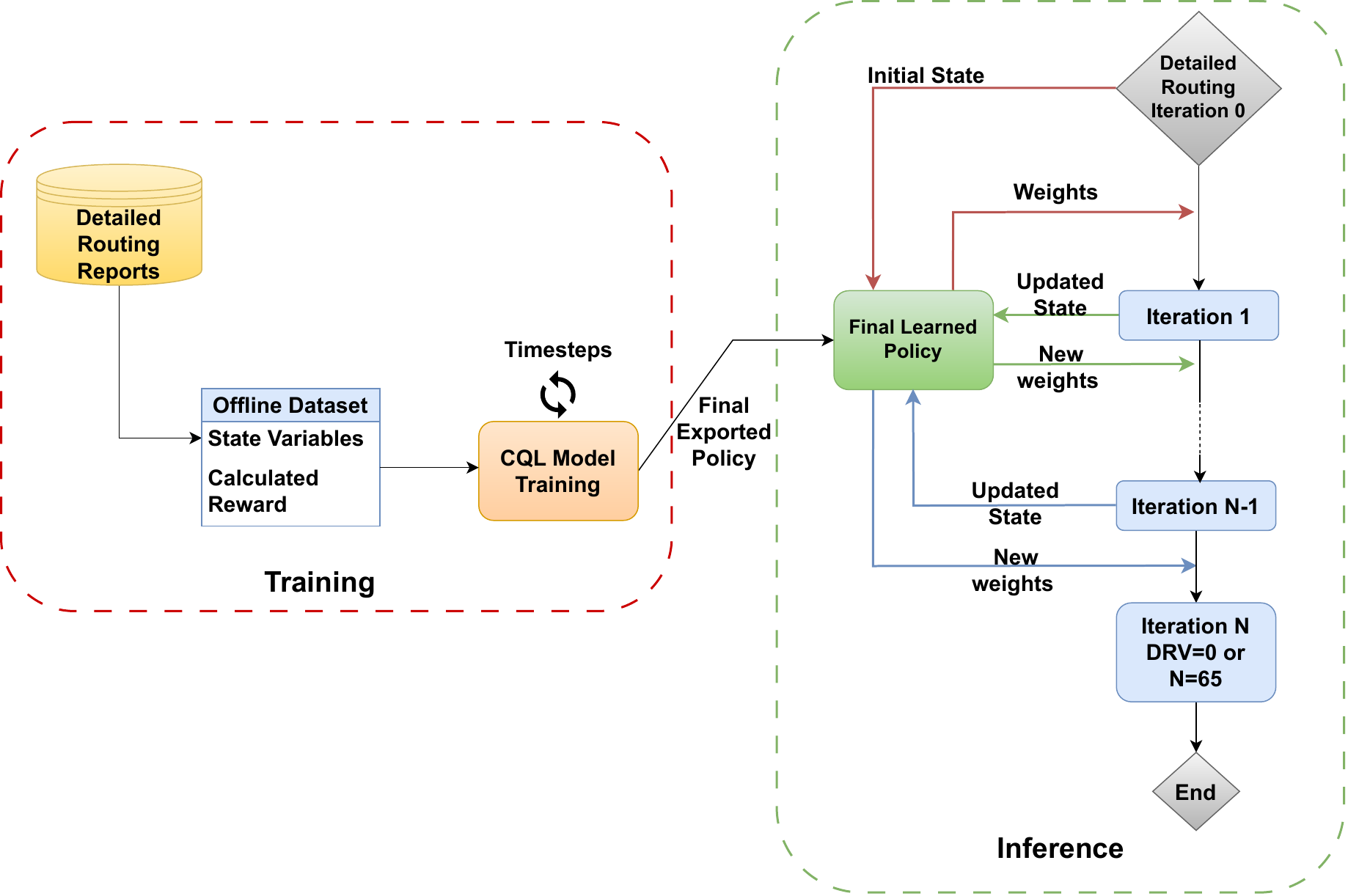}
    \caption{Training and inference flow for conservative Q-Learning model}\vspace{-4mm}
    \label{fig:example}
\end{figure}

\subsection{State Representations and Reward Functions}
For the reinforcement learning model to effectively train and predict optimal weights for iteration $n$, a state representation from iteration $n-1$ should convey meaningful design characteristics and dynamic routing progress information to the model.
In addition to the variables collected in Section~\ref{sec:DataGeneration}, the variables derived from the data that were included in the state representation are:
\begin{itemize}
    \item Average density of DRVs per routing region
    \item Rate of DRV reduction across iterations
\end{itemize}

Feature vectors were normalized using standard scaler from training data statistics. The reward function was clipped to 1st-99th percentiles of training data and normalized using tanh before training to help maintain stability with higher learning rates during training. The variables in the reward function in the order of importance are:
\begin{itemize}
    \item DRV Improvement Bonus (highest importance)
    \item Iteration Penalty (highest importance)
    \item Convergence Bonus  (highest importance)
    \item Stuck Penalty (low importance)
\end{itemize}

We deliberately exclude fine-grained spatial coordinates of violations to keep the state representation simple and generalizable. Our proposal aims at balancing model complexity while establishing a novel framework for applying reinforcement learning to a highly capable router.

\subsection{Model Tuning}

\begin{table}
\begin{center}
\caption{CQL Hyperparameter Values}\vspace{-3mm}
\label{tunetable}
    \begin{tabular}{|c|c|} \hline 
        \textbf{Hyperparameter} &  \textbf{Value} \\ \hline 
        \texttt{Actor Learning Rate} & $1.0 \times 10^{-3}$\\ \hline 
        \texttt{Critic Learning Rate} & $4.0 \times 10^{-3}$\\ \hline 
        \texttt{Conservative Weight}& $8.0 \times 10^{-1}$\\ \hline 
        \texttt{Batch Size} & 128\\ \hline 
        \texttt{Initial Temperature} & $1.43298 \times 10^{-1}$ \\ \hline 
        \texttt{Temperature LR} & 0.0\\ \hline 
        \texttt{Tau (Polyak $\tau$)} & $1.97555 \times 10^{-3}$ \\ \hline 
    \end{tabular}
    \vspace{-6mm}
\end{center}    
\end{table}

Our model was tuned to provide faster convergence in good weight configurations without succumbing to overestimation or out-of-distribution shifts. Using the hyperparameters suggested by CQL~\cite{cql} and optimizing with d3rlpy~\cite{d3rlpy}, we arrived at the values shown in Table~\ref{tunetable}.
To prevent policy collapse during training, we implement two early stopping mechanisms monitored every epoch:

\noindent
\textbf{Critic Loss Explosion Detection}: The training pipeline monitors critic loss for exponential growth, which indicates Q-function approximation failure. When the critic network fails to track the actor's rapidly changing policy updates, cascading instability occurs where the actor receives increasingly unreliable gradient signals. We implement a threshold of 100x increase to accommodate the transient overshooting behavior that can occur during early training when actor and critic networks bootstrap off each other (critic's Bellman targets) before stabilizing for the remainder of the training.

\noindent
\textbf{Action Difference Monitoring}: The d3rlpy framework's action\_diff metric computes $\text{action\_diff} = \frac{1}{N}\sum_{i=1}^{N}\|\pi_\theta(s_i) - a_i\|^2$, where $\pi_\theta(s_i)$ represents the learned policy's action and $a_i$ represents the action from the dataset for state $s_i$. While some deviation is expected for policy improvement, excessive divergence (\textgreater{}1.0) indicates out-of-distribution behavior where generalization fails. Since actions are normalized to [0,1] during training, action\_diff is particularly sensitive and represents a substantial deviation in the original action space. 

\section{Experimental Results}
\label{sec:ExperimentalResults}
\subsection{Methodology}
For our experimental evaluation, we implement a comprehensive pipeline to train, deploy, and validate our reinforcement learning approach.
For deployment, we developed a real-time inference pipeline using TorchScript and LibTorch to integrate the saved feature scaler and trained policy directly into OpenROAD's C++ detailed router. This implementation extracts the relevant state variables at the end of each routing iteration and feeds them into the trained RL policy to predict optimal weight configurations for the subsequent iteration. To evaluate performance, we conducted comparative experiments between the baseline (default) weight configurations and RL-predicted weights across 29 standard and benchmark circuits, with 10 being unseen test benchmarks from ISPD19.
Our benchmarks were run on an AMD EPYC 9275F CPU @ 4.1 GHz with 768 GB of DDR5 RAM. All training was performed on this platform, as the data collection time far outweighed the model training time.
The runtime for each benchmark was averaged over 10 runs for each configuration (default and RL-guided).
Runtimes for our approach include policy inference overhead, which was measured to be about 2s each run.

\subsection{Model Training}

\begin{figure}
    \centering
    \includegraphics[width=0.9\linewidth]{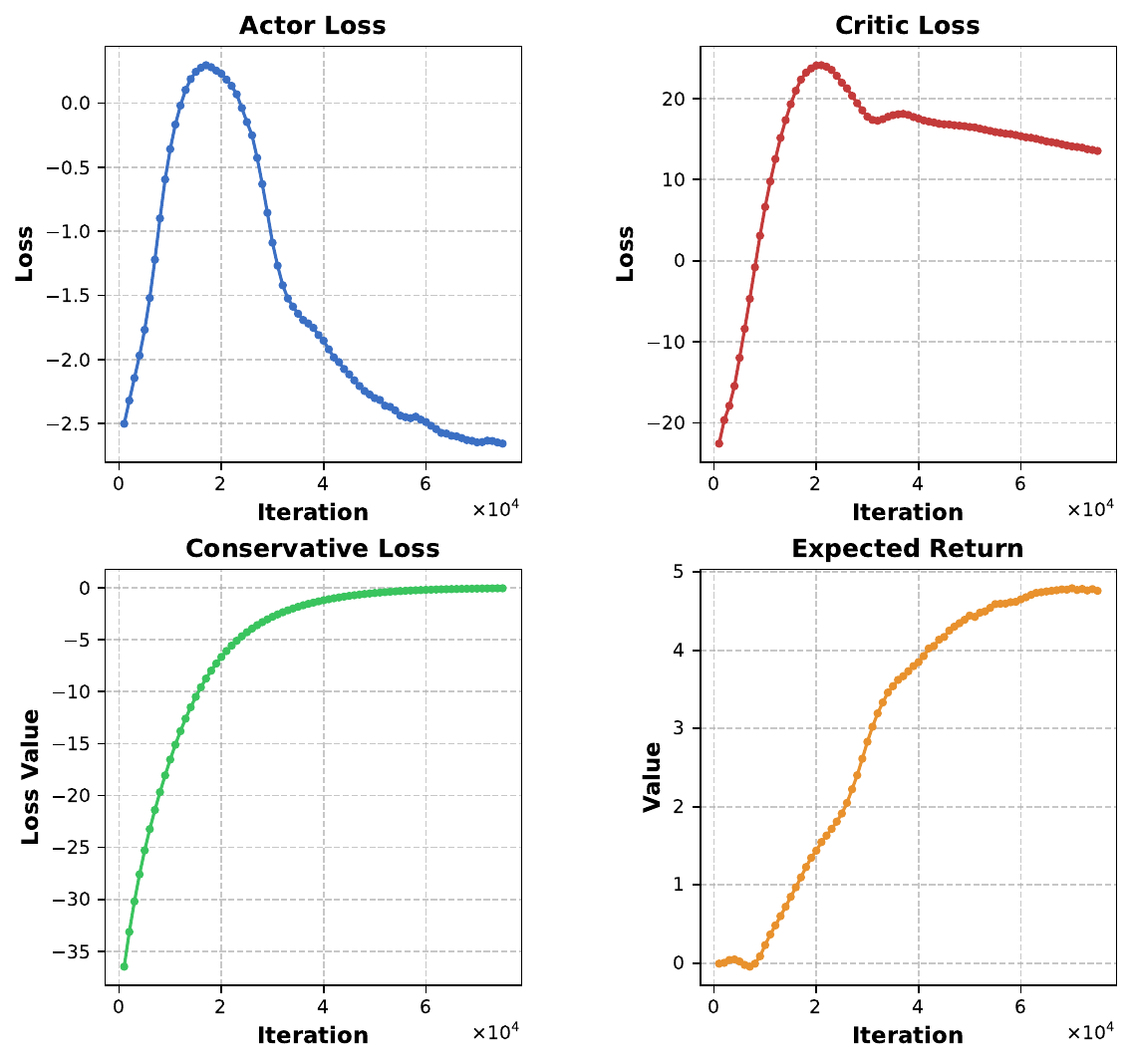}\vspace{-2mm}
    \caption{CQL Training Progress} \vspace{-4mm}
    \label{fig:trainingProgress}
\end{figure}

Fig.~\ref{fig:trainingProgress} shows four standard training metrics from the CQL model.
In our RL model,
both the critic and actor curves overshoot before settling to lower values. This initial spike validates our 
high threshold for early stopping,
allowing model to recover from transient instabilities. 
Early on, the Q-network chases shifting targets and the policy chases the evolving Q-estimates.
With continued training, both losses steadily descend due to the 
convergence of both the actor and critic networks.
Similarly, the conservative penalty starts very negative to strongly discourage out-of-distribution actions, but gradually rises to zero as the policy learns align with the training data \cite{cql}.
The Expected Return plot tracks the agent's estimate of the cumulative reward from the initial Q (first routing state), thus the smooth monotonic rise confirms that the learned policy is maximizing the expected return \cite{rlparam}. We monitor the Initial State Value Estimation metric to determine convergence and allow the training to continue until this metric plateaus with variance \textless{}0.01 over 3 epochs, indicating stable value  function approximation.
The training duration shown in Fig. \ref{fig:trainingProgress} corresponds to the point where the expected return was observed to consistently approach convergence across multiple 
training runs without triggering the early stopping criteria.


\subsection{Benchmark Performance}

\begin{table*}
 \begin{center}  
\caption{Performance on OpenROAD Design Suite (Seen Data)}\vspace{-3mm}
\label{tab1}
    \begin{tabular}{|r|r|r|r|r|r|r|r|r|r|r|r|} \hline 
          \textbf{Design}&  \multicolumn{3}{|c|}{\textbf{Iterations}} & \multicolumn{3}{|c|}{\textbf{Runtime (s)}} &  \multicolumn{2}{|c|}{\textbf{DRVs}} & \multicolumn{3}{|c|}{\textbf{Wirelength (um)}}\\ \hline 
          & \textbf{\textit{Base}} & \textbf{\textit{Ours}} & \textbf{\textit{Min.}}
          & \textbf{\textit{Base}} & \textbf{\textit{Ours}} & \textbf{\textit{Diff}}
          & \textbf{\textit{Base}} & \textbf{\textit{Ours}}
          & \textbf{\textit{Base}} & \textbf{\textit{Ours}} &\textbf{\textit{Diff}} \\ \hline 
 \texttt{aes}        & \s6 & \textbf{5} & 4 &  55 &  46&16.36\%& 0 & 0 & \s309750 &\s312744& -0.97\% \\ \hline 
 \texttt{ariane136}  & \s6 & \textbf{4} & 4 & 147& 143& 2.72\%& 0 & 0 & \s8017226&\s8038308&-0.26\%\\ \hline 
 \texttt{bp\_be}     &  15 &  \textbf{11} & 5& 115&  56&51.30\%& 0 & 0 & \s3028121&\s3037392&-0.31\%\\ \hline 
 \texttt{bp\_fe}     &  13 & \textbf{7}& 5 & 80&  43&46.25\%& 0 & 0 & \s2352172&\s2359893&-0.33\%\\ \hline 
 \texttt{bp\_multi}  &  11 & \textbf{5}& 4& 132&  110&16.67\%& 0 & 0 & \s4711286&\s4729301&-0.38\%\\ \hline 
 \texttt{gcd}        &   4 & \textbf{3}& 2 &   3&   4&-33.33\%& 0 & 0 & \s4786&\s4827&-0.86\%\\ \hline 
 \texttt{ibex}       &   5 & \textbf{4}& 4 &  27 &  28 &-3.70\% & 0 & 0 & \s330625&\s332567&-0.59\% \\ \hline 
 \texttt{jpeg}       & \s5 & \textbf{4} & 4 &  39&  37&5.13\%& 0 & 0 & 1163854&\s1169491&-0.48\%\\ \hline 
 \texttt{swerv}      & \s5 & \textbf{4} & 4 &  94&  96&-2.13\%& 0 & 0 & \s2959953&\s2973861&-0.47\%\\ \hline 
 \texttt{tinyRocket} & \s5 & \textbf{4} & 4&  21 &  20&4.76\% & 0 & 0 & \s723782&\s726170&-0.33\% \\ \hline
 \textbf{Total}      & --  & --  & -- &  713 &  \textbf{583} & \textbf{18.23\%} & 0 & 0 & 23601555
&23684554
&-0.35\%
\\ \hline
 \end{tabular}
 \vspace{-5mm}
\end{center}
\end{table*}

\newcommand{\fixme}{{\color{red}?}}

\begin{table*}
\caption{Performance on ISPD19 (Unseen Test Data)}\vspace{-5mm}
\begin{center}
\begin{tabular}{|r|r|r|r|r|r|r|r|r|r|r|}
\hline
\textbf{Design} &
  \multicolumn{2}{|c|}{\textbf{Iterations}} &
  \multicolumn{3}{|c|}{\textbf{Runtime (s)}} &
  \multicolumn{2}{|c|}{\textbf{DRVs}} &
  \multicolumn{3}{|c|}{\textbf{Wirelength (\si{\micro\meter})}} \\ \hline 
&
  \textbf{\textit{Base}} & \textbf{\textit{Ours}} &
  \textbf{\textit{Base}} & \textbf{\textit{Ours}} & \textbf{\textit{Diff}} &
  \textbf{\textit{Base}} & \textbf{\textit{Ours}} &
  \textbf{\textit{Base}} & \textbf{\textit{Ours}}  & \textbf{\textit{Diff}}\\ \hline
 \texttt{ispd19\_test1}  & \s\textbf{5} & \s\textbf{5} & \s\s\textbf{43}& \s\s46& -6.98\% & \s0 & \s0 & \s\s\s63293 & \s\s\s63694 & -0.63\%\\ \hline
 \texttt{ispd19\_test2}  & \s\textbf{7} & \s\textbf{7} &  \s255 &  \s\textbf{155} & \s39.22\% & \s0 & \s0 & \s2480245 & \s2473440 & 0.27\%\\ \hline 
 \texttt{ispd19\_test3}  & \s\textbf{6} & \s\textbf{6} &  \s106 &  \s\textbf{104} & \s1.89\% & \s0 & \s0 & \s\s\s82647 & \s\s\s82683 & -0.04\%\\ \hline 
 \texttt{ispd19\_test4}  &  16          &  \textbf{11}&  1336 &  \s\textbf{443}& 66.84\%& \s0 & \s0 & \s6000839 & \s6013707& -0.21\%\\ \hline 
 \texttt{ispd19\_test6}  & \s6          & \s\textbf{5} &  \s289 &  \s\textbf{262} & \s9.34\% & \s0 & \s0 & \s6536852 & \s6542048 & -0.08\%\\ \hline 
 \texttt{ispd19\_test7}  & \s\textbf{5} & \s\textbf{5} &  \s\textbf{315} &  \s317 & -0.63\% & \s0 & \s0 & 12181108 & 12189586 & -0.07\%\\ \hline 
 \texttt{ispd19\_test8}  & \s\textbf{5} & \s\textbf{5} &  \s458 &  \s\textbf{455} & \s0.66\% & \s0 & \s0 & 18733926 & 18733268 & 0.00\%\\ \hline
 \texttt{ispd19\_test9}  & \s\textbf{5} & \s\textbf{5} &  \s\textbf{758} &  \s772 & -1.85\% & \s0 & \s0 & 28347604 & 28354901 & -0.03\%\\ \hline
 \texttt{ispd19\_test10} &  \textbf{65} &  \textbf{65} &   4538 &   \textbf{2736} & 39.71\% &  24 &  5 & 28032422 & 28052084 & -0.07\%\\ \hline
 \textbf{Total} &           -- &           -- &   8098 &   \textbf{5200} & \textbf{35.79\%} &  24 &  5& 102458936 & 102504963 & -0.04\%\\ \hline
\end{tabular}
\label{tab2}\vspace{-5mm}
\end{center}
\end{table*}

\begin{figure}
    \centering
    \includegraphics[width=1\linewidth]{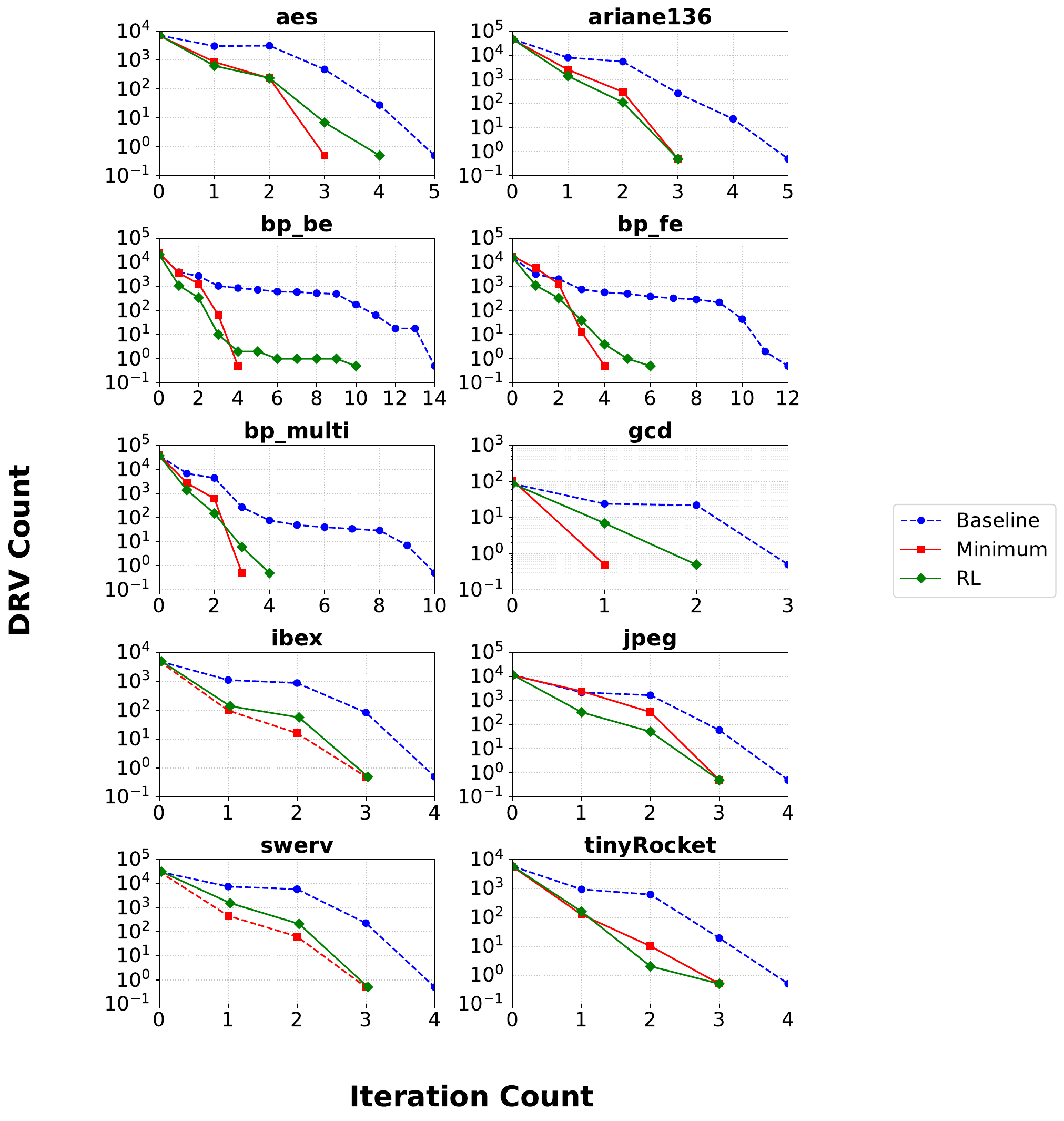}\vspace{-3mm}
    \caption{DRV convergence comparison for the OpenROAD Design Suite}\vspace{-1mm}
    \label{fig:drvConvergence}
\end{figure}

Our experimental results are summarized in Tables~\ref{tab1} and \ref{tab2}. Table~\ref{tab1} shows the results across 1) the baseline OpenROAD detailed router, 2) our work using the RL model, and 3) the fastest (min.) converging model seen in the generated data. Each of the designs in this set are designs which the model has trained on, and therefore shows how well the model is able to perform on previously seen data.
Table~\ref{tab1} shows the results across 1) the baseline router and 2) our RL model. 
Because the training data included ISPD18 benchmarks, we report here only the ISPD19 benchmarks to represent completely unseen designs (including a new technology node).
Across both seen and unseen data, our model is successful in reducing the number of iterations. Our model finishes in the same or fewer iterations as the baseline.
For runtime, our model is successful in decreasing the average runtime across both sets of data as well.
It should be noted that in several cases, the number of iterations does not decrease, but the runtime does. This typically occurs when the policy reduces DRVs earlier in the trajectory, which lowers the amount of work required in later iterations. Figure~\ref{fig:drvConvergence} shows DRV traces for our model and the baseline and illustrates this effect. 

The policy impact is also benchmark dependent. Speedups are most pronounced on larger or more congested designs where each iteration is expensive and early DRV reduction compounds over time. In contrast, several smaller designs such as \texttt{gcd} and \texttt{ibex} already converge quickly under the baseline. In these cases, there is limited routing work to reduce, so the inference overhead per iteration can dominate the total runtime which yields negligible gain or a slight slowdown. We include the inference overhead in all reported runtimes. 

This effect is visible on \texttt{ispd19\_test9}, where runtime increases despite similar convergence behavior. Upon inspection, the routing difficulty of this benchmark appears relatively low for the baseline and the remaining runtime is dominated by processing a large number of simple nets rather than resolving difficult violations. Since the policy does not substantially change the routing effort in this regime, the constant inference overhead can outweigh the small savings. On the other hand, \texttt{ispd19\_test10} exhibits higher congestion and benefits substantially from our method. The terminal DRV count is also reduced in this case, although both our model and the baseline reach the maximum iteration cap and finish with non-zero DRVs.
For wirelength, on the seen data there is a small penalty of \textless{}1\% on average and therefore the impact is minimal. We did not include wirelength in the reward function, and therefore it was not considered as an optimization target when selecting the weights.

\subsection{Weight Analysis}
\begin{figure}
    \centering
    \includegraphics[width=0.9\linewidth]{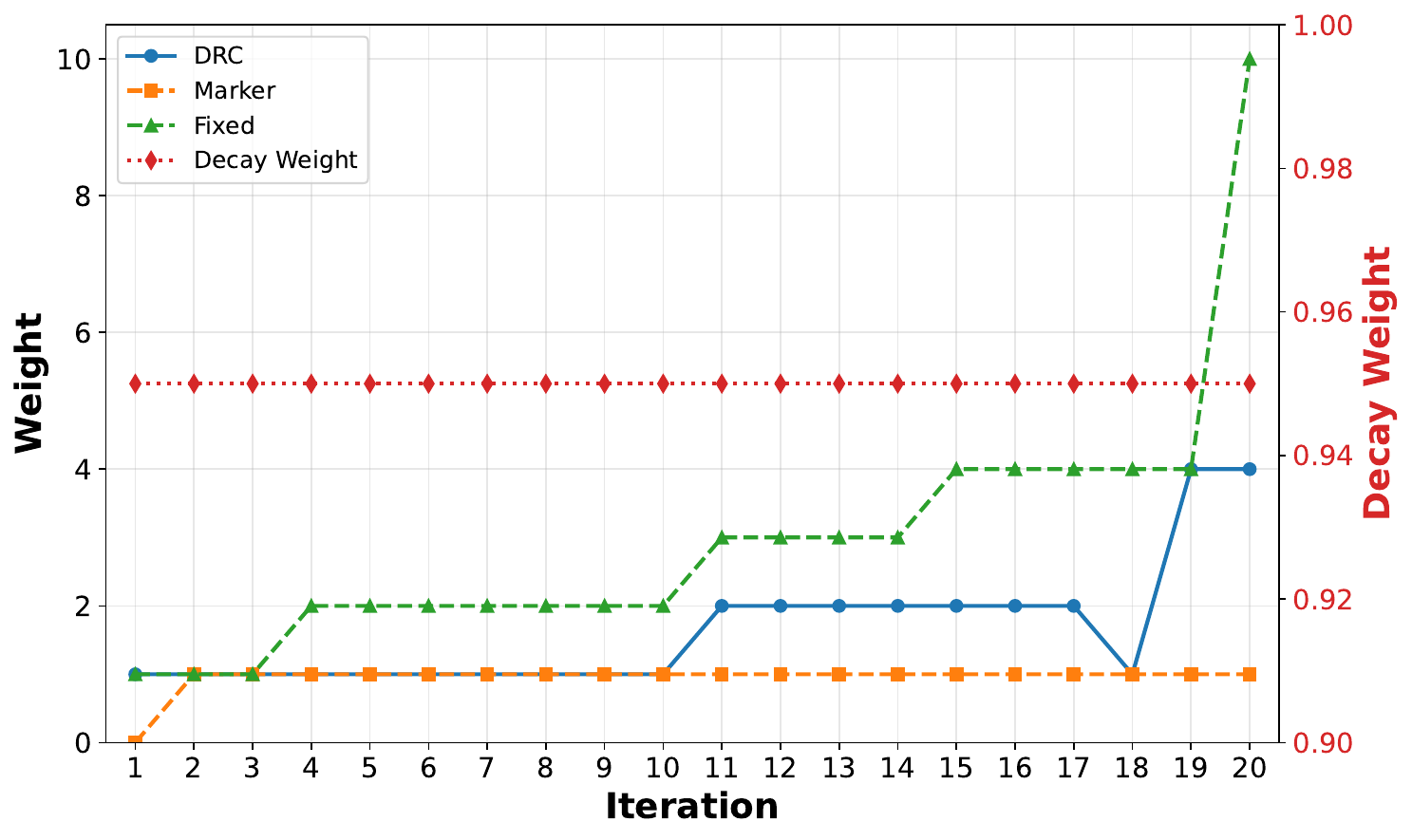}\vspace{-1mm}
    \caption{Baseline weight configuration for all designs. Truncated after iteration 20 for brevity. Designs will stop early if number of DRVs goes to 0.}\vspace{-3mm}
    \label{fig:baselineWeights}
\end{figure}

\begin{figure}
   \vspace{-6mm}
   \centering
   \includegraphics[width=1\linewidth]{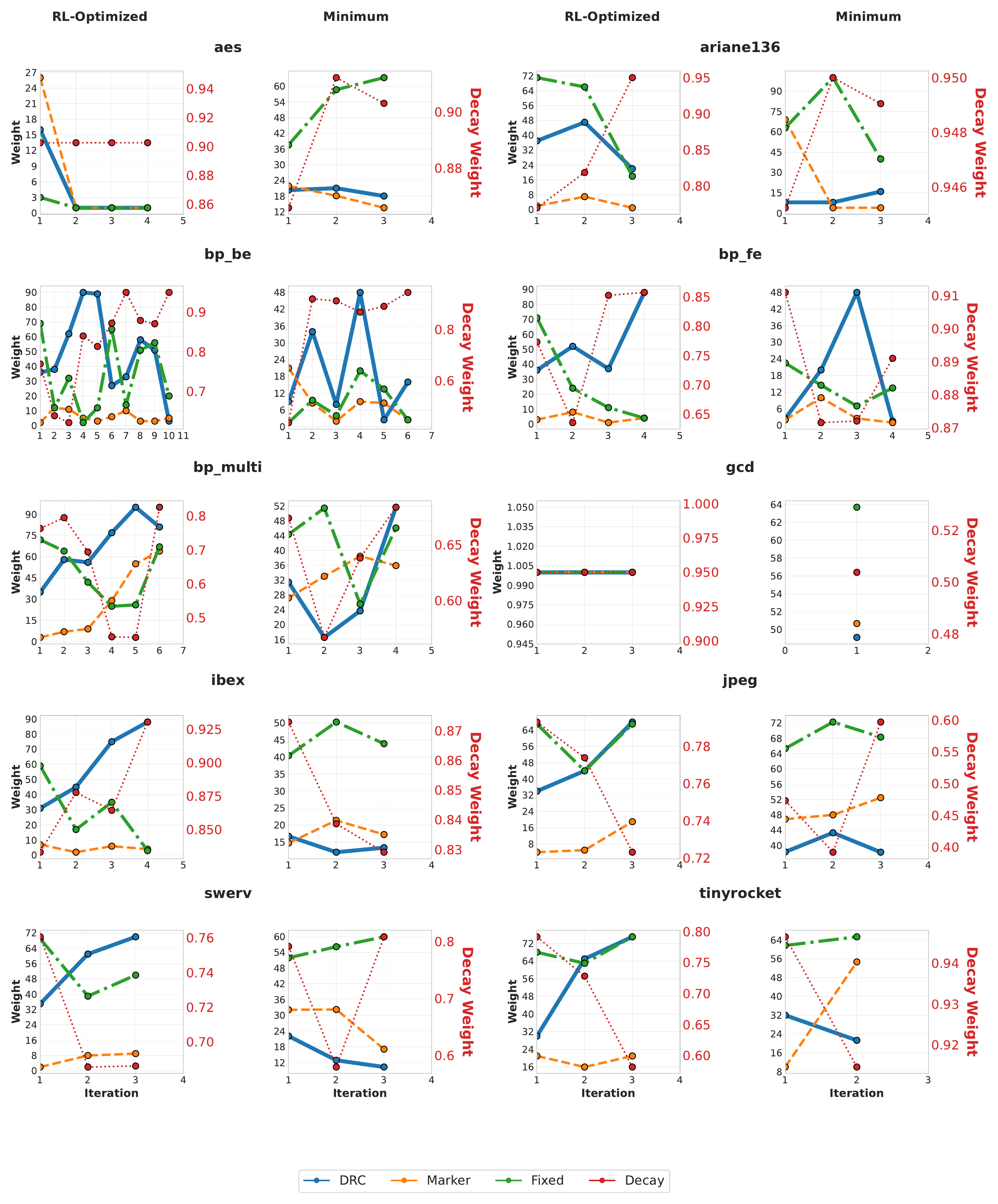}
   \caption{Weight selections by our RL model vs. weights used in the fastest converging sequence in the training data.}
   \label{fig:weightGrid}\vspace{-6mm}
\end{figure}

Fig.~\ref{fig:baselineWeights} shows the full schedule of weights for the baseline implementation up until iteration 20 (truncated for brevity), and Fig.~\ref{fig:weightGrid} shows the weights for both the ones inferred at runtime by our RL model and the weights for the fastest converging sequence in the training data. The weight trajectories in Fig.~\ref{fig:baselineWeights} indicate that our model learns generalized strategies by exploring its own path and not simply replicating the optima from the training data, thus balancing the bias-variance tradeoff for robustness on unseen designs. This also suggests there may be more than one weight trajectory leading to identical results since cost functions work in relativity.

Comparing the weight trajectories as a whole, it is clear that the weights required to minimize the iteration count can vary largely based on the design. Our model was able to identify features that enabled it to dynamically select better weights than the baseline version.

\section{Conclusion}
\label{sec:Conclusion}

This paper presented a detailed router solution which leverages reinforcement learning to enable rapid convergence in detailed routing. By training a conservative Q-learning model to infer A* search weights and minimize iterations, we were able to reduce the iterations by 5\% on average and up to 31\% on unseen designs in the ISPD19 benchmark. This translates to a runtime speedup of \avgSpeedup{} on average and up to \maxSpeedup{}. We found that this learning generalizes to other technologies, as ISPD19 has technology nodes that the model has not seen before, yet it still demonstrated significant improvement.

\section*{Acknowledgments}
We thank Eugene Vinitsky for helpful early conversations and Saik Anam Siam for early guidance on practical machine learning directions for detailed routing.
This work was partially funded by a Google charitable gift.

\bibliographystyle{IEEEtran}
\bibliography{refs}

@string{iccad = "Proceedings of the IEEE/ACM International Conference on
	Computer-Aided Design"}

@string{dac = "Proceedings of the ACM/IEEE Design Automation Conference"}

@string{aspdac = "Proceedings of the Asia-South Pacific Design Automation
	Conference"}

@string{ispd = "Proceedings of the International Symposium on Physical Design"}

@string{iccad = "Proceedings of the IEEE/ACM International Conference on Computer-Aided Design"}

@string{aspdac = "Proceedings of the Asia-South Pacific Design Automation Conference"}

@string{iccad = " Proc. ICCAD"}

@string{dac = " Proc. DAC"}

@string{aspdac = " Proc. ASP-DAC"}

@string{ispd = " Proc. ISPD"}

@inproceedings{ajayi2019openroad-gomactech,
  title={{OpenROAD: Toward a Self-Driving, Open-Source Digital Layout Implementation Tool Chain}},
  author={Ajayi, Tutu and Blaauw, David and Chan, Tuck-Boon and Cheng, Chung-Kuan and Chhabria, Vidya A. and Choo, David K. and Coltella, Matteo and Dobre, Sorin and Dreslinski, Ronald G. and Fogaça, Mateus and Hashemi, Soheil and Hosny, Abdelrahman and Kahng, Andrew B. and Kim, Minsoo and Li, Jiajia and Liang, Zhaoxin and Mallappa, Uday and Penzes, Paul and Pradipta, Geraldo and Reda, Sherief and Rovinski, Austin and Samadi, Kambiz and Sapatnekar, Sachin S. and Saul, Lawrence and Sechen, Carl and Srinivas, Vaishnav and Swartz, William and Sylvester, Dennis and Urquhart, David and Wang, Lutong and Woo, Mingyu and Xu, Bangqi},
  booktitle = {Proc. GOMATECH},
  year = {2019}
}

@INPROCEEDINGS{ajayi2019openroad,
  author={Ajayi, Tutu and Chhabria, Vidya A. and Fogaça, Mateus and Hashemi, Soheil and Hosny, Abdelrahman and Kahng, Andrew B. and Kim, Minsoo and Lee, Jeongsup and Mallappa, Uday and Neseem, Marina and Pradipta, Geraldo and Reda, Sherief and Saligane, Mehdi and Sapatnekar, Sachin S. and Sechen, Carl and Shalan, Mohamed and Swartz, William and Wang, Lutong and Wang, Zhehong and Woo, Mingyu and Xu, Bangqi},
  booktitle=dac, 
  title={{INVITED: Toward an Open-Source Digital Flow: First Learnings from the OpenROAD Project}}, 
  year={2019},
  volume={},
  number={},
  doi={}
}

@inproceedings{rovinski2020bridging,
 author = {Rovinski, Austin and Ajayi, Tutu and Kim, Minsoo and Wang, Guanru and Saligane, Mehdi},
 title = {{Bridging Academic Open-Source EDA to Real-World Usability}},
 booktitle = iccad,
 pages = {1--7},
 month = {November},
 year = {2020}
}

@article{mishra2021enhancing,
  title={{Enhancing Accuracy of Deep Learning Algorithms by Training with Low-Discrepancy Sequences}},
  author={Mishra, Siddhartha and Rusch, T Konstantin},
  journal={SIAM Journal on Numerical Analysis},
  volume={59},
  number={3},
  pages={1811--1834},
  year={2021},
  publisher={SIAM}
}

@inproceedings{shi2023provably,
  title={{Provably Efficient Offline Reinforcement Learning with Perturbed Data Sources}},
  author={Shi, Chengshuai and Xiong, Wei and Shen, Cong and Yang, Jing},
  booktitle={International Conference on Machine Learning},
  pages={31353--31388},
  year={2023},
  organization={PMLR}
}

@article{kahng2020tritonroute,
  title={{TritonRoute: The Open-Source Detailed Router}},
  author={Kahng, Andrew B and Wang, Lutong and Xu, Bangqi},
  journal={IEEE Transactions on Computer-Aided Design of Integrated Circuits and Systems},
  volume={40},
  number={3},
  pages={547--559},
  year={2020},
  publisher={IEEE}
}

@inproceedings{kahng2018tritonroute,
  title={{TritonRoute: An Initial Detailed Router for Advanced VLSI Technologies}},
  author={Kahng, Andrew B and Wang, Lutong and Xu, Bangqi},
  booktitle={2018 IEEE/ACM International Conference on Computer-Aided Design (ICCAD)},
  pages={1--8},
  year={2018},
  organization={IEEE}
}

@ARTICLE{lee1961algorithm,
  author={Lee, C. Y.},
  journal={IRE Transactions on Electronic Computers}, 
  title={{An Algorithm for Path Connections and Its Applications}}, 
  year={1961},
  volume={EC-10},
  number={3},
  pages={346-365},
  doi={10.1109/TEC.1961.5219222}}

@INPROCEEDINGS{perturbation,
  author={Liu, Qianmei and Kuang, Yufei and Wang, Jie},
  booktitle={2024 International Joint Conference on Neural Networks (IJCNN)}, 
  title={{Robust Deep Reinforcement Learning with Adaptive Adversarial Perturbations in Action Space}}, 
  year={2024},
  volume={},
  number={},
  pages={1-8},
  doi={10.1109/IJCNN60899.2024.10651543}}

@inproceedings{sobol,
	doi = {https://doi.org/10.26868/25222708.2011.1590},
	url = {https://publications.ibpsa.org/conference/paper/?id=bs2011_1590},
	year = {2011},
	month = {November},
	publisher = {IBPSA},
	author = {Sebastian Burhenne and Dirk Jacob and   Gregor P. Henze},
	title  = {Sampling Based on Sobol{$'$} Sequences for Monte Carlo Techniques Applied to Building Simulations},
	booktitle = {Proceedings of Building Simulation 2011: 12th Conference of IBPSA},
	volume  = {12},
	isbn = {},
	address  = {Sydney, Australia},
	series  = {Building Simulation},
	pages = {1816--1823},
	issn = {2522-2708},
	Organisation = {IBPSA},
	Editors = {}
}

@inproceedings{cql,
 author = {Kumar, Aviral and Zhou, Aurick and Tucker, George and Levine, Sergey},
 booktitle = {Advances in Neural Information Processing Systems},
 editor = {H. Larochelle and M. Ranzato and R. Hadsell and M.F. Balcan and H. Lin},
 pages = {1179--1191},
 publisher = {Curran Associates, Inc.},
 title = {{Conservative Q-Learning for Offline Reinforcement Learning}},
 volume = {33},
 year = {2020}
}

@misc{offlinerl2,
      title={{Offline Reinforcement Learning: Tutorial, Review, and Perspectives on Open Problems}}, 
      author={Sergey Levine and Aviral Kumar and George Tucker and Justin Fu},
      year={2020},
      eprint={2005.01643},
      archivePrefix={arXiv},
      primaryClass={cs.LG},
}

@ARTICLE{rlbook,
  author={Sutton, R.S. and Barto, A.G.},
  journal={IEEE Transactions on Neural Networks}, 
  title={{Reinforcement Learning: An Introduction}}, 
  year={1998},
  volume={9},
  number={5},
  pages={1054-1054},
  doi={10.1109/TNN.1998.712192}}

@INPROCEEDINGS{10,
  author={Zeng, Wei and Davoodi, Azadeh and Topaloglu, Rasit Onur},
  booktitle={2020 Design, Automation \& Test in Europe Conference \& Exhibition (DATE)}, 
  title={{Explainable DRC Hotspot Prediction with Random Forest and SHAP Tree Explainer}}, 
  year={2020},
  volume={},
  number={},
  pages={1151-1156},
  doi={10.23919/DATE48585.2020.9116488}}

@ARTICLE{11,
  author={Park, Hyunbum and Baek, Kyeonghyeon and Kim, Suwan and Choi, Kyumyung and Kim, Taewhan},
  journal={IEEE Transactions on Computer-Aided Design of Integrated Circuits and Systems}, 
  title={{Pin Accessibility and Routing Congestion Aware DRC Hotspot Prediction for Designs in Advanced Technology Nodes With Consolidated Practical Applicability and Sustainability}}, 
  year={2024},
  volume={43},
  number={12},
  pages={4786-4799},
  doi={10.1109/TCAD.2024.3405894}}

@inproceedings{12,
    author = {Liang, Rongjian and Xiang, Hua and Pandey, Diwesh and Reddy, Lakshmi and Ramji, Shyam and Nam, Gi-Joon and Hu, Jiang},
    title = {{DRC Hotspot Prediction at Sub-10nm Process Nodes Using Customized Convolutional Network}},
    year = {2020},
    isbn = {9781450370912},
    publisher = {Association for Computing Machinery},
    address = {New York, NY, USA},
    doi = {10.1145/3372780.3375560},
    booktitle = {Proceedings of the 2020 International Symposium on Physical Design},
    pages = {135–142}
}

@inproceedings{13,
author = {Chen, Hao and Hsu, Kai-Chieh and Turner, Walker J. and Wei, Po-Hsuan and Zhu, Keren and Pan, David Z. and Ren, Haoxing},
title = {{Reinforcement Learning Guided Detailed Routing for Custom Circuits}},
year = {2023},
isbn = {9781450399784},
publisher = {Association for Computing Machinery},
address = {New York, NY, USA},
doi = {10.1145/3569052.3571874},
booktitle = {Proceedings of the 2023 International Symposium on Physical Design},
pages = {26–34},
numpages = {9},
location = {Virtual Event, USA},
series = {ISPD '23}
}

@article{bidir,
author = {Kaindl, Hermann and Kainz, Gerhard},
title = {{Bidirectional Heuristic Search Reconsidered}},
year = {1997},
issue_date = {July 1997},
publisher = {AI Access Foundation},
address = {El Segundo, CA, USA},
volume = {7},
number = {1},
issn = {1076-9757},
journal = {J. Artif. Int. Res.},
month = dec,
pages = {283–317},
numpages = {35}
}

@inproceedings{arnold,
author = {Arnold, Michael H. and Scott, Walter S.},
title = {{An Interactive Maze Router with Hints}},
year = {1988},
isbn = {0818688645},
publisher = {IEEE Computer Society Press},
address = {Washington, DC, USA},
booktitle = {Proceedings of the 25th ACM/IEEE Design Automation Conference},
pages = {672–676},
numpages = {5},
location = {Atlantic City, New Jersey, USA},
series = {DAC '88}
}

@inproceedings{han2015,
author = {Han, Kwangsoo and Kahng, Andrew B. and Lee, Hyein},
title = {{Evaluation of BEOL Design Rule Impacts Using an Optimal ILP-Based Detailed Router}},
year = {2015},
isbn = {9781450335201},
publisher = {Association for Computing Machinery},
address = {New York, NY, USA},
doi = {10.1145/2744769.2744839},
booktitle = {Proceedings of the 52nd Annual Design Automation Conference},
articleno = {68},
numpages = {6},
location = {San Francisco, California},
series = {DAC '15}
}

@inproceedings{nieberg2011,
author = {Nieberg, Tim},
title = {Gridless pin access in detailed routing},
year = {2011},
isbn = {9781450306362},
publisher = {Association for Computing Machinery},
address = {New York, NY, USA},
doi = {10.1145/2024724.2024763},
booktitle = {Proceedings of the 48th Design Automation Conference},
pages = {170–175},
numpages = {6},
location = {San Diego, California},
series = {DAC '11}
}

@article{ding2017,
author = {Ding, Yixiao and Chu, Chris and Mak, Wai-Kei},
title = {{Self-Aligned Double Patterning Lithography Aware Detailed Routing With Color Preassignment}},
year = {2017},
issue_date = {Aug. 2017},
publisher = {IEEE Press},
volume = {36},
number = {8},
issn = {0278-0070},
doi = {10.1109/TCAD.2016.2622625},
month = aug,
pages = {1381–1394},
numpages = {14}
}

@INPROCEEDINGS{liu2014,
  author={Iou-Jen Liu and Shao-Yun Fang and Yao-Wen Chang},
  booktitle={2014 51st ACM/EDAC/IEEE Design Automation Conference (DAC)}, 
  title={{Overlay-Aware Detailed Routing for Self-Aligned Double Patterning Lithography Using the Cut Process}}, 
  year={2014},
  volume={},
  number={},
  pages={1-6},
  doi={10.1109/DAC.2014.6881377}}

@article{chang2013,
author = {Ahrens, Markus and Gester, Michael and Klewinghaus, Niko and Muller, Dirk and Peyer, Sven and Schulte, Christian and Tellez, Gustavo},
title = {{Detailed Routing Algorithms for Advanced Technology Nodes}},
year = {2015},
issue_date = {April 2015},
publisher = {IEEE Press},
volume = {34},
number = {4},
issn = {0278-0070},
doi = {10.1109/TCAD.2014.2385755},
journal = {Trans. Comp.-Aided Des. Integ. Cir. Sys.},
month = apr,
pages = {563–576},
numpages = {14}
}

@inproceedings{hightower1969solution,
author = {Hightower, David W.},
title = {{A Solution to Line-Routing Problems on the Continuous Plane}},
year = {1969},
isbn = {9781450379298},
publisher = {Association for Computing Machinery},
address = {New York, NY, USA},
url = {https://doi.org/10.1145/800260.809014},
doi = {10.1145/800260.809014},
booktitle = {Proceedings of the 6th Annual Design Automation Conference},
pages = {1–24},
numpages = {24},
series = {DAC '69}
}

@misc{rlparam,
      title={{Hyperparameter Selection for Offline Reinforcement Learning}}, 
      author={Tom Le Paine and Cosmin Paduraru and Andrea Michi and Caglar Gulcehre and Konrad Zolna and Alexander Novikov and Ziyu Wang and Nando de Freitas},
      year={2020},
      eprint={2007.09055},
      archivePrefix={arXiv},
      primaryClass={cs.LG},
      url={https://arxiv.org/abs/2007.09055}, 
}

@misc{d3rlpy,
      title={{d3rlpy: An Offline Deep Reinforcement Learning Library}}, 
      author={Takuma Seno and Michita Imai},
      year={2022},
      eprint={2111.03788},
      archivePrefix={arXiv},
      primaryClass={cs.LG},
      url={https://arxiv.org/abs/2111.03788}, 
}

@inproceedings{chen2020detailed,
  author = {Chen, Gengjie and Pui, Chak-Wa and Li, Haocheng and Chen, Jingsong and Jiang, Bentian and Young, Evangeline F. Y.},
  title = {{Detailed Routing by Sparse Grid Graph and Minimum-Area-Captured Path Search}},
  year = {2019},
  isbn = {9781450360074},
  publisher = {Association for Computing Machinery},
  address = {New York, NY, USA},
  url = {https://doi.org/10.1145/3287624.3287678},
  doi = {10.1145/3287624.3287678},
  booktitle = {Proceedings of the 24th Asia and South Pacific Design Automation Conference},
  pages = {754–760},
  numpages = {7},
  location = {Tokyo, Japan},
  series = {ASPDAC '19}
}

@inproceedings{mantik2018ispd,
author = {Mantik, Stefanus and Posser, Gracieli and Chow, Wing-Kai and Ding, Yixiao and Liu, Wen-Hao},
title = {{ISPD 2018 Initial Detailed Routing Contest and Benchmarks}},
year = {2018},
isbn = {9781450356268},
publisher = {Association for Computing Machinery},
address = {New York, NY, USA},
url = {https://doi.org/10.1145/3177540.3177562},
doi = {10.1145/3177540.3177562},
booktitle = {Proceedings of the 2018 International Symposium on Physical Design},
pages = {140–143},
numpages = {4},
location = {Monterey, California, USA},
series = {ISPD '18}
}

@inproceedings{liu2019ispd,
author = {Liu, Wen-Hao and Mantik, Stefanus and Chow, Wing-Kai and Ding, Yixiao and Farshidi, Amin and Posser, Gracieli},
title = {{ISPD 2019 Initial Detailed Routing Contest and Benchmark with Advanced Routing Rules}},
year = {2019},
isbn = {9781450362535},
publisher = {Association for Computing Machinery},
address = {New York, NY, USA},
url = {https://doi.org/10.1145/3299902.3311067},
doi = {10.1145/3299902.3311067},
booktitle = {Proceedings of the 2019 International Symposium on Physical Design},
pages = {147–151},
numpages = {5},
location = {San Francisco, CA, USA},
series = {ISPD '19}
}

\end{document}